\documentclass{iopart}
\usepackage{iopams}
\usepackage{epsfig}
\usepackage{graphicx}
\usepackage{version}
\usepackage{amssymb}
\usepackage{float}

\newcommand{\dd}{\textrm{d}}
\newcommand{\ee}{\rme}

\newcommand{\gammat}{{\tilde \gamma}}
\newcommand{\Dt}{{\tilde D}}
\newcommand{\D}{{\Delta}}
\newcommand{\Dg}{{\Delta \gamma}}

\newcommand{\vd}{{\dot v}}
\newcommand{\hP}{{\hat P}}
\newcommand{\hrho}{{\hat \rho}}

\newcommand{\p}{\partial}
\graphicspath{{figures/}}

\begin{document}
\title{Work fluctuations for a Brownian particle between two thermostats} 

\date{\today}
\author{Paolo Visco}
\address{Laboratoire de Physique Th\'eorique (CNRS UMR 8627), B\^atiment
  210, Universit\'e Paris-Sud, 91405 Orsay cedex, France}
\address{Laboratoire de Physique Th\'eorique
et Mod\`eles Statistiques (CNRS
UMR 8626), B\^atiment 100, Universit\'e Paris-Sud, 91405 Orsay cedex, France}
\ead{Paolo.Visco@th.u-psud.fr}

\begin{abstract}
We explicitly determine the large deviation function of the energy
flow of a Brownian particle coupled to two heat baths at different
temperatures. This toy model, initially introduced by Derrida and
Brunet \cite{derridabrunet}, allows not only to sort out the influence
of initial conditions on large deviation functions but also to
pinpoint various restrictions bearing upon the range of validity of
the Fluctuation Relation.
\end{abstract}

\submitto{Journal of Statistical Mechanics: theory and experiment}
\maketitle

\begin{figure}
\begin{centering}
  \includegraphics[width=0.65 \textwidth,clip=true]{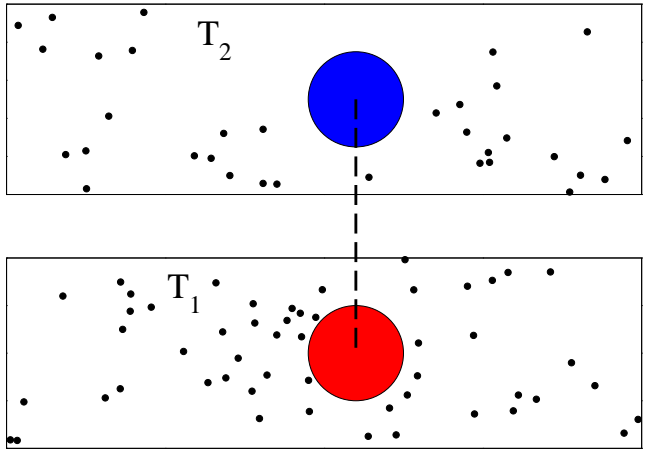}
  \caption{\label{ratchet} (color online) Sketch of a model, inspired
    by \cite{vandenbroeck} described by the Langevin equation
    (\ref{langevin}). The two particles are constrained to move
    together along the horizontal axis, and are in two heat baths at
    different temperatures.}
\end{centering}
\end{figure}

The recent upsurge of interest in nonequilibrium statistical mechanics lies in
the discovery of simple yet generic results, embodied by the Fluctuation
Relation initially brought forth by Evans, Cohen and Morriss
\cite{evanscohenmorriss}, and later formalized into a theorem by Gallavotti
and Cohen \cite{gallavotticohen}. This Fluctuation Relation states that in the
nonequilibrium steady-state of a (chaotic, Anosov) dynamical system, the
temporal large deviation function of an appropriately defined entropy current
verifies a particular symmetry property under time reversal. In practice the
latter entropy current consists of a heat or particle flux and could be
accessed experimentally in some situations~\cite{experiments}.  Before going
into the specifics of that relation, we mention that a Markov dynamics analog
of the Gallavotti-Cohen theorem was then found by Kurchan \cite{kurchan} and
Lebowitz and Spohn \cite{lebowitzspohn}. More recently, other contributions on
the subject have been put forward by Van Zon and Cohen \cite{vanzoncohen}, and
by Bonetto {\it et al.} \cite{bonetto}, where an extension of the Fluctuation
Relation to quantities other than the entropy flux is proposed.\\
The purpose of the present report is to present a simple model for which
nonequilibrium fluctuations can be investigated precisely at finite times. We
will show that a seemingly innocuous initial condition dependence restricts
the domain of validity of the Fluctuation Relation, even though the latter
applies to asymptotic times.\\
In order to illustrate our point, we exploit a toy model introduced by Derrida
and Brunet in a recent pedagogical account \cite{derridabrunet}. After having
described this model, we will recall the standard statement of the Fluctuation
Relation, and then we shall show that some initial condition effects, that
intuition suggests to discard, can lead to an actual failure of the
Fluctuation Relation.\\
We take a particle in contact with two heat baths imposing distinct
temperatures $T_1$ and $T_2$, whose velocity $v(t)$ evolves according to the
Langevin equation
\begin{equation}
\label{langevin}
m \vd = - (\gamma_1 + \gamma_2) v + \xi_1(t) + \xi_2(t)\,\,,
\end{equation} 
with $\langle \xi_i \rangle = 0$ and
\begin{equation}
\langle \xi_i(t) \xi_j(t') \rangle=2 \gamma_i
T_i \delta_{ij} \delta(t-t') = 2 D_i \delta_{ij} \delta(t-t')\,\,,
\end{equation}
where the Boltzmann's constant has been taken equal to one. Such an equation
can also describe the dynamics of apparently different models, for example,
the dynamics of a hard rod connected at both ends to two heat baths (as a
slight variation of the model introduced in \cite{vandenbroeck}, and is
sketched in Fig \ref{ratchet}).  We also note that another model, consisting
of two overdamped brownian particles interacting with a harmonic potential
\cite{marin,fred}, can be exactly mapped on Eq. (\ref{langevin}).  Finally, we
quote that other analogies in term of electric circuits, in the spirit of
\cite{ciliberto} can also be described by a Stochastic differential equation
of the same type.
Strictly speaking, the dynamics of the particle is an equilibrium one,
since in Eq. (\ref{langevin}) the sum of the Gaussian noises is
another Gaussian noise with viscosity $\gamma=\gamma_1 + \gamma_2$ and
noise strength $D=D_1 + D_2 $. The velocity probability distribution
function (pdf) of the particle is hence a Gaussian with zero-mean and
variance $T_0=D/\gamma$. Nevertheless, for our purpose, we distinguish
the thermostats and we focus on the heat flowing from one
  thermostat (say for example thermostat 1) to the particle over
a time duration $t$. This quantity is exactly the total work performed
by the thermostat, and it reads:
\begin{equation}
\label{Qs}
Q_i=\int_0^t \dd \tau \, v(\tau) (\xi_i(\tau) - \gamma_i v(\tau))\,\,, 
\end{equation}
where $i=1,2$. The pdf of this integrated injected power $P(Q_1,t)$ is
the central object of our note. The Fluctuation Relation, applied
without care, states without restriction that for the fluctuating time
averaged flux $q =Q_1/t$, the large deviation function (ldf) $\pi(q)$
defined as
\begin{equation}
\pi(q)=\lim_{t\to\infty}\frac{1}{t}\ln P(q\;t,t)\,\,
\end{equation}
verifies the symmetry property
\begin{equation}\label{FR2}
\pi(q)-\pi(-q)=\epsilon q \;.
\end{equation}
This is the Fluctuation Relation (FR). In Eq.\,(\ref{FR2}), the constant
$\epsilon=\frac{1}{T_2}-\frac{1}{T_1}$ plays the role of an external field
driving the heat flux. In the following, we will also consider a restriction
of the relation (\ref{FR2}) to a finite interval (i.e. for $|q| < q^*$, where
$q^*$ is a positive constant), in analogy with deterministic systems
\cite{gallavotti,bonetto} and stochastic systems \cite{vanzoncohen}.  We will
refer to this last relation as ``Extended Fluctuation Relation'' (EFR), from
the terminology of \cite{vanzoncohen}. Usually the Fluctuation Relation is
supposed to be verified by the entropy flux ${\cal S}=Q_1/T_1+Q_2/T_2$, which
differs from $\epsilon \, Q_1$ only by a total time difference, and hence,
following \cite{bonetto}, $Q_1$ should, at least, verify an Extended
Fluctuation Relation.

\begin{figure}[t]
\begin{centering}
 \includegraphics[clip=true,width=0.6 \textwidth]{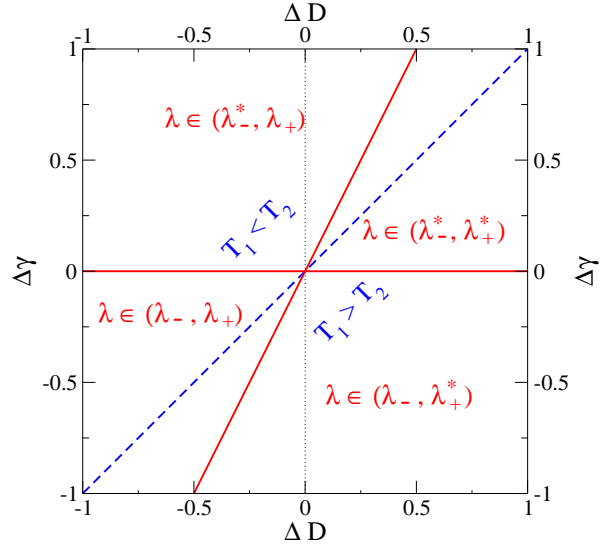}
  \caption{\label{phasediagram} (color online) Diagram for the extra
  cuts in the real $\lambda$ axis. The solid lines separate the
  regions of the parameter space $(\Dg, \D D)$ for which $\hP
  (\lambda)$ has different domains of definition in the $\lambda$-real
  axis. The dashed line is the line where $T_1=T_2$.}
\end{centering}
\end{figure}

We now wish to explicitly determine $P(Q_1,t)$ at all time and
investigate the properties of the large deviation function. The
techniques used are similar to those employed by Farago
\cite{farago}. We first introduce the joint probability $\rho
(v,Q_1,t)$ that the particle has a velocity $v$ and a given value of
$Q_1$ at a time $t$. The pdf $P(Q_1,t)$ is obtained by integrating
$\rho$ over the velocity $v$. We also define the Laplace Transform
\begin{equation}
\hrho (v, \lambda ,t) =\int \dd Q_1 \, \ee^{-\lambda Q_1}
\rho(v,Q_1,t) \,\,,
\end{equation}
which verifies: (i) $\hrho(v,0,t)=f(v,t)$ with $f(v,t)$ the velocity
pdf of the brownian particle, and (ii) $\hP(\lambda,t) = \int \dd v
\hrho(v,\lambda,t)$, with $\hP(\lambda, t)$ the Laplace Transform of
$P(Q_1,t)$. The pdf $\hrho (v,\lambda,t)$ verifies a Fokker-Planck
Equation (FPE):
\begin{equation}
\label{FPE}
\frac{\p}{\p t} \hrho(v,\lambda,t) = L_{\lambda} \hrho(v,\lambda,t)\,\,,
\end{equation}
where \footnote{A detailed derivation of this operator for a slightly
  different system is given in: P. Visco, A. Puglisi, A. Barrat, E.
  Trizac and F. van Wijland, submitted to J. Stat. Phys, {\tt
  cond-mat/0509487}.}
\begin{equation}
L_{\lambda}=D \frac{\p^2}{\p v^2}
+ (\gamma + 2 \lambda D_1) \frac{\p}{\p v} v  
+(D_1 \lambda^2 + \gamma_1 \lambda) v^2 - D_1 \lambda\,\,.
\end{equation}
In order to get the solution of the above FPE we first note that the
eigenvalue equation
\begin{equation}
L_{\lambda} f_n(v,\lambda)=\mu_n(\lambda) f_n(v,\lambda)
\end{equation}
is solved by:
\begin{equation}
\mu_n (\lambda) = \frac{\gamma}{2} (1- (1+2 n) \eta)\,\,,
\end{equation}
\begin{equation}
f_n(v, \lambda) = \frac{\ee^{- \frac{v^2}{2 T(\lambda)}}}
{\sqrt{2 \pi T(\lambda)}} 
H_n \left( \frac{v}{\sqrt{2 T^*(\lambda)}} \right)/\sqrt{2^n n!}\,\,,
\end{equation}
where:
\begin{equation}
\eta=\sqrt{1+ \frac{4 \lambda}{\gamma^2} (\gamma_2 D_1 - \gamma_1 D_2 
- \lambda  D_1 D_2)}\,\,,
\end{equation}
\begin{equation}
T(\lambda)= \frac{2 D}{\gamma (1 + \eta) + 2 \lambda D_1}\,,
\quad T^*(\lambda)=T_0/\eta\,\,,
\end{equation}
and where $H_n(x)$ denotes the Hermite polynomial of order $n$. We
remark that the largest eigenvalue $\mu_0(\lambda)$ presents the
symmetry
\begin{equation}
\label{FR3}
\mu_0(\lambda) = \mu_0 (\epsilon - \lambda)\,\,,
\end{equation}
with $\epsilon = 1/T_2 - 1/T_1$. Should $\pi(q)$ be the Legendre
Transform of $\mu_0(\lambda)$, this last relation would exactly yield
(\ref{FR2}) \cite{lebowitzspohn}. The solution of (\ref{FPE}) for a
given initial condition is hence easily obtained as:
\begin{equation}
  \hrho(v,\lambda,t|v_0)=\sum_{n=0}^{\infty} \ee^{\mu_n (\lambda) t}
  C_n(v,\lambda|v_0) f_n(v,\lambda)\,\,,
\end{equation}
where $C_n (v, \lambda|v_0)$ is the projection of the $n$-th
eigenfunction onto the initial state, which has been chosen to be a
Dirac function centered in $v=v_0$. In order to simplify the notations
we will set all the scales to unity (i.e. $\gamma=D=T_0=1$), and
introduce
\begin{equation}
\gammat_1=\frac{\gamma_1}{\gamma}=\frac{1+\Dg}{2}\,,\quad
\gammat_2=\frac{\gamma_2}{\gamma}=\frac{1-\Dg}{2}\,\,,
\end{equation}
\begin{equation}
\Dt_1=\frac{D_1}{D}=\frac{1+\D D}{2}\,,\quad
\Dt_2=\frac{D_2}{D}=\frac{1-\D D}{2}\,\,,
\end{equation}
where both $\Dg$ and $\D D$ take values between $-1$ and $1$.  Thus,
when $T_1=T_2$ one has that $\Dg = \D D$. Finally, integrating
$\hrho$ over the velocity, one obtains the expression of $\hP
(\lambda,t)$ for a given initial velocity $v_0$:
\begin{equation}
\fl
\label{pilambdav0}
\hP(\lambda,t | v_0)=\ee^{\frac{t}{2}} \left(\cosh (\eta t) + (\lambda
(1+ \D D) + 1) \frac{\sinh (\eta t)}{\eta} \right)^{-\frac{1}{2}} \\
\exp \left(\frac{v_0^2}{2} \left(\frac{\lambda (1+\Dg + (1+\D D)
\lambda)} {1+\lambda (1 + \D D) + \eta \coth (\eta t)} \right) \right)
\,\,.
\end{equation}
The long time behavior of $\hP(\lambda,t)$ is clearly dominated by
$\hP(\lambda,t) \sim \ee^{\mu_0(\lambda) t}$, with $\mu_0(\lambda)$
the largest eigenvalue of $L_{\lambda}$. This result was already found
in \cite{derridabrunet}. The expression of $\mu_0$ presents two cuts
in the real axis for $\lambda > \lambda_+$ and $\lambda < \lambda_-$,
with:
\begin{equation}
\lambda_{\pm}= \frac{\D D - \Dg}{1- \D D^2} \pm \sqrt{\frac{1-2 \D D \Dg 
+ \Dg^2}{(1- \D D^2)^2}}\,\,.
\end{equation}
Nonetheless, it is possible to see that the subleading prefactor
entering in expression (\ref{pilambdav0}) presents an extra cut for
$\lambda < \lambda_-^*$, where, in the infinite time limit,
\begin{equation}
\lambda_-^*=-\frac{1+\Dg}{1+\D D}\,\,.
\end{equation}
Note that this cut exists only for $\Dg > 0$, and implies that the
right tail of the large deviation function of $Q_1$ will present an
exponential decay, with a slope different than the one predicted by
the Legendre Transform of $\mu_0$. The quantitative details on how
this extra cut will affect $\pi(q)$ will be given later. Let us now
give some comments on this issue. The energy balance equation for this
particular system can be written as:
\begin{equation}
\D E(t) = E(t)- E(0)=Q_1+Q_2\,\,,
\end{equation}
where $Q_2$ denotes the work performed by the thermostat at
temperature $T_2$ on the particle, and is defined by Eq (\ref{Qs}).
Physical intuition would suggest that for very long times, since the
$Q_i$'s are extensive in time while the ``boundary term'' $\D E$ is
not, the latter does not come into play when determining the large
deviation of, say, $Q_1$, and in particular that $Q_1$ and $-Q_2$ have
the same large deviation functions. However, in several examples
\cite{farago,farago2,vanzoncohen,tracer,bonetto} it has been noted
that, if the boundary term distribution has exponentially (or slower)
decreasing tails then, even in the infinite time limit, it can have a
fluctuation of order $t$, and hence is no more negligible. Clearly in
our case, since the velocity pdf is Gaussian, the energy has
exponential tails. If the initial condition $v_0$ is fixed, then $\D
E$ has only a right exponential tail, which may affect the right tail
of the large deviation function of $Q_1$. This is precisely the reason
for the presence of our extra cut. Besides, if the initial condition
is sampled on the stationary velocity pdf, the boundary term $\D E$
will have two exponential tails, and therefore two extra cuts may
appear in the $\lambda$ complex plane. When the initial condition is
sampled in the stationary state, one finds:
\begin{equation}
\fl \hP(\lambda)=\int \dd v_0 \, \ee^{-\frac{v_0^2}{2}}
\hP(\lambda|v_0)= \ee^{\frac{t}{2}} \times \\ \left(\cosh(\eta t) +
\frac{\sinh(\eta t)}{\eta} (1-\lambda (\Dg + \D D (\lambda -1) +
\lambda)) \right)^{\frac{1}{2}}\,.
\end{equation}
In the infinite time limit this last expression still presents the
$\lambda_-^*$ extra cut for $\Dg >0$. Moreover, a new cut appears for
$\D D> \Dg/2$ and for $\lambda > \lambda_+^*$ with
\begin{equation}
\lambda_+^*= \frac{1+ 2 \D D - \Dg}{1+ \D D}\,\,.
\end{equation}
These cuts are schematized in a diagram in figure \ref{phasediagram}.

\begin{figure}
\begin{centering}
  \includegraphics[width=.6 \textwidth,clip=true]{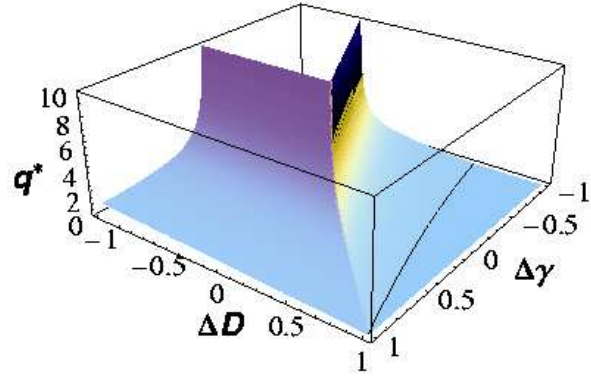}
  \caption{\label{qstar} (color online) Three dimensional plot
    of $q^*$, as a function of the system's parameters. In the region
    in which $q^*=\infty$ the FR is verified. The solid line
    delimitates the region in which $q^*=0$, where the EFR fails.}
\end{centering}
\end{figure}

\begin{figure}
\begin{centering}
  \includegraphics[clip=true,width=0.6
  \textwidth]{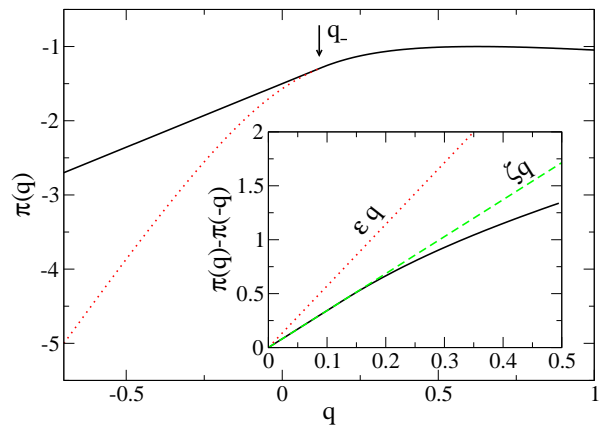}
  \caption{\label{pidiq} Plot of the large deviation function $\pi(q)$
    for $\Delta D = 0.75$ and $\Delta \gamma=-0.5$. The (red) dotted
    line shows $\pi_c(q)$, for which the Fluctuation Relation
    holds. The inset shows the plot of $\pi(q)-\pi(-q)$. The (red)
    dotted line is a straight line of slope $\epsilon$, while the
    (green) dashed line has a slope $\zeta$ (see text for details).}
\end{centering}
\end{figure}

The Laplace Transform can be safely inverted using the Legendre
Transform only in the region in which:
\begin{equation}
\max (\lambda_-,\lambda_-^*) < \lambda < \min
(\lambda_+,\lambda_+^*)\,\,.
\end{equation}
The Legendre Transform gives hence in this region:
\begin{equation}
\pi_c(q)= - \frac{1- \D D^2 + 2 q (\Dg -\D D) + \sqrt{(1+\Dg^2 - 2 \Dg
\D D) (1+4 q^2 - \D D^2)}}{2 (1- \D D^2)}\,\,.
\end{equation}
If the bounds of $\hP(\lambda)$ are (on the real axis) $\lambda_+$ and
$\lambda_-$ the above expression is valid for any $q$, and the
Fluctuation Relation holds. Besides, if the left (right) cut begins at
$\lambda_-^*$ ($\lambda_+^*$), the Legendre Transform of
$\mu_0(\lambda)$ is valid only for $q<q_+=\mu_0'(\lambda_-^*)$
($q>q_-=\mu_0'(\lambda_+^*)$). In this case the expression of $\pi(q)$
outside the interval $(q_-,q_+)$ is a straight line $\pi (q)= \alpha +
\beta q$, and the coefficients $\alpha$ and $\beta$ can be obtained
using that both $\pi(q)$ and $\pi'(q)$ must be continuous functions of
$q$ (this is not evident for the derivative $\pi'(q)$, but it is
effectively the case in \cite{farago}, where essentially the same
model has been worked out). For example in the region $\Dg>0$ and $\D
D > \Dg/2$ one has: 
\begin{figure}[t]
\begin{centering}
  \includegraphics[clip=true,width=0.6
  \textwidth]{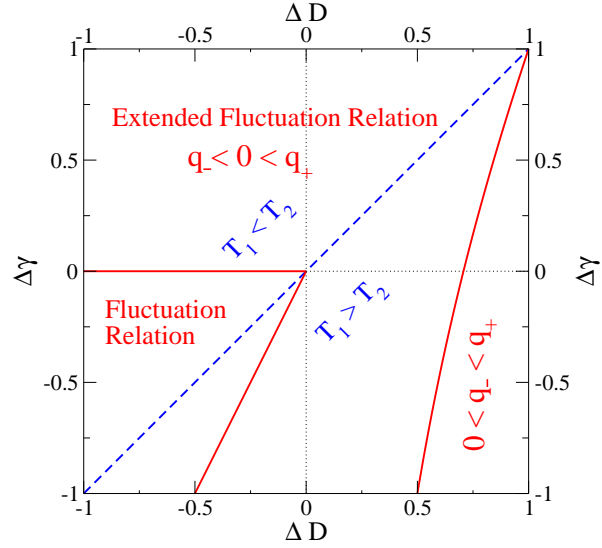}
  \caption{\label{qdiagram}(color online) Diagram schematizing the domain of
    validity of the Fluctuation Relation and of the Extended
    Fluctuation Relation (see text for details).}
\end{centering}
\end{figure}

\begin{equation}
\pi(q)= \cases{ \pi_l(q) \qquad q<q_-\,\,\\ \pi_c (q) \qquad q_-< q <
q_+\,\,\\ \pi_r(q) \qquad q > q_+\,\,,}
\end{equation}
with
\begin{equation}
q_-=\frac{\D D}{2}-\frac{1}{2 ( 2 \D D- \Dg)}\,,\quad q_+=\frac{1}{2
\Dg}- \frac{\D D}{2}\,\,,
\end{equation}
\begin{equation}
\pi_l(q)=-\frac{(1+ 2 \D D - \Dg ) (1 + \D D - 2 q)}{2 (1 + \D D)}\,\,,
\end{equation}
\begin{equation}
\pi_r(q)=-\frac{(1+ \Dg)(1+ \D D + 2 q)}{2 (1+\D D)}\,\,.
\end{equation}
The behavior of $\pi(q)$ in the others regions can be obtained in a similar
fashion. If an extra cut starts at $\lambda_-^*$, then for $q>q_+$ the ldf
$\pi(q)$ is equal to $\pi_r(q)$. Analogously, if an extra cut begins at
$\lambda_+^*$, then for $q<q_-$ one has $\pi(q)=\pi_l(q)$. In all the other
cases $\pi(q)=\pi_c(q)$. The Fluctuation Relation (\ref{FR2}) is only
satisfied in the case for which $\pi(q)=\pi_c(q)$. The Extended Fluctuation
Relation is satisfied in the cases where $q_-<0<q_+$ (see the diagram in Fig
\ref{qdiagram}). In this case $q^*$ is the minimum between $|q_-|$ and
$|q_+|$, and is plotted in Fig \ref{qstar}. Surprisingly it also happens that,
in a in a given region of the parameters' space (i.e. when $\Delta \gamma < 2
\Delta D - 1/\Delta D$) one has that $0<q_-<q_+$. In this case $q^*=0$ and
both the FR and the EFR break down. In particular it is found, for small
values of $q$, that:
\begin{equation}
\pi(q)-\pi(-q)=\pi_l(q)-\pi_l(-q)= \zeta q\,, \qquad |q|<q_-\,\,
\end{equation} 
with $\zeta =4/T_0-2/T_1$ (an illustration is given in Fig
\ref{pidiq}). \\

We now turn on some remarks concerning the importance of initial conditions.
If the two temperatures $T_1$ and $T_2$ are equal, it is clear that the
average flux $\langle Q_1 \rangle =0$, with $\pi(q)=\pi(-q)$. Surprisingly, if
the ldf $\pi(q|v_0)$ is measured for $\Dg>0$ (and $T_1=T_2$), with a fixed
initial condition $v_0$, one would see that $\pi(q|v_0) \ne \pi(-q|v_0)$
(because of the right extra cut), which would lead to the wrong impression
that there is a nonzero flux from one thermostat to the other, even if they
are at the same temperature. Of course, if $\pi(q)$ is measured sampling the
initial conditions in the stationary state, a new extra cut appears, restoring
the symmetry $\pi(q)=\pi(-q)$. This can be understood with the following
remark. The right tail of $\pi(q|v_0)$ is somehow dominated by the events with
an ``energetic final condition'' (cf.  also \cite{farago}). The bias enforced
by the fixed initial condition (even in the case where $v_0=\langle v
\rangle=0$), forbids the occurrence of ``energetic initial conditions'', which
would balance (on average) the energetic final conditions, and leads then to
the appearance of an unphysical flux.

Our results answer some issues raised in the recent developments of
the theory of nonequilibrium fluctuations. In particular, we have
provided the confirmation that if two quantities, both extensive in
time (as e.g. $Q_1 (t)$ and $Q_2 (t)$), differ one from the other by a
``boundary term'' (as $\D E (t)$), whose pdf has tails with
exponential decay, they do not have the same large deviation
function. This also leads to a small restriction to the validity of
the very general result obtained in \cite{kurchan, lebowitzspohn}.
Furthermore, our results clearly show that, even if the Fluctuation
Relation (\ref{FR2}) is broken, the largest eigenvalue
$\mu_0(\lambda)$ always displays the symmetry property analogous to
(\ref{FR3}). While the latter result is obviously mathematically
robust, this is also the most relevant physics-wise, since for small
fluctuations around the average and for small nonequilibrium drive, it
leads to the well-known fluctuation-dissipation theorem.

\ack I thank A. Puglisi, E. Trizac and F. van Wijland for stimulating
discussions and for a careful reading of this manuscript. This work
was supported by the Agence Nationale de la Recherche through
an ACI.\\

\end{document}